\documentclass[twocolumn,showpacs,preprintnumbers,amsmath,amssymb, pr]{revtex4}

\usepackage{graphicx}
\usepackage{dcolumn}
\usepackage{bm}
\usepackage{epsfig}

\begin{document}

\title{Three-Dimensional Phase-Kink State in Thick Stack of Josephson Junctions \\
and Terahertz Radiation}

\author{Xiao Hu\(^{1,2,3}\) and Shizeng Lin\(^{1,2}\)}

\affiliation{\(^{1}\)WPI Center for Materials Nanoarchitectonics,
 National Institute for Materials Science, Tsukuba 305-0047, Japan\\
\(^{2}\)Graduate School of Pure and Applied Sciences, University of
Tsukuba, Tsukuba 305-8571, Japan\\
\(^{3}\)Japan Science and Technology Agency, 4-1-8 Honcho,
Kawaguchi, Saitama 332-0012, Japan}
\date{\today}

\begin{abstract}
The dynamics of superconductivity phase in thick stack of Josephson
junctions with strong inductive coupling, such as the one realized
in layered high-$T_c$ cuprates and possibly the recently discovered
FeAs-based superconductors, is investigated under a $c$-axis bias
voltage and in the absence of an external magnetic field. The kink
state found previously by the present authors is extended to three
dimensions for both rectangular and cylindrical geometries. The IV
characteristics are calculated and the distributions of
electromagnetic field inside the samples are clarified.  The
solution for a cylindrical mesa exhibits a higher resonating
frequency than that of a square mesa with the same linear size by a
factor of $\sim 2.4$. More importantly, from the radius dependence
of the resonance frequency for the cylinder geometry it is possible
to confirm directly the kink state, and thus to reveal the mechanism
of the strong radiation discovered in recent experiments.
\end{abstract}


\maketitle

\section{Introduction}
Stimulated by the recent progress in radiation of terahertz (THz)
 electromagnetic waves using a mesa of
 $\rm{Bi_2Sr_2CaCu_2O_{8+\delta}}$ (BSCCO) single
crystal \cite{Ozyuzer07,Kadowaki08}, we investigated the dynamics of
superconductivity phase of the intrinsic Josephson junctions with
strong inductive couplings. A new dynamics state was found in the
absence of an external magnetic field, in which the gauge invariant
phase difference in each junction is characterized by static
$\pm\pi$ phase kinks stacked periodically along the $c$ axis, in
addition to a phase term evolving linearly with time according to
the ac Josephson relation and the plasma term
\cite{Lin08PRL,Koshelev08arXiv}. The phase kink permits pumping
large dc energy into the plasma oscillation via the ac Josephson
effect, and enhances strong radiations of THz electromagnetic waves.
This state is the only one known so far which counts for the
following important features of the coherent radiations observed in
Refs.~\cite{Ozyuzer07,Kadowaki08}: (1) the frequency is determined
by the cavity mode of the mesa sample, (2) the frequency and voltage
satisfy the ac Josephson relation, and (3) the radiation takes place
in a sharp regime of voltage.

For simplicity, a two dimensional (2D) solution was worked out
explicitly, presuming a state uniform along the in-plane direction
of the long edge of the mesa \cite{Lin08PRL}. This solution
corresponds to the (1,0) cavity mode. In the present work, we show
that the kink state can be extended to 3D, where the phase kink runs
in the $ab$ planes, and the solution is labeled by the (1,1) cavity
mode. The kink state can also exist in samples of the cylinder
geometry.

We first derive a set of equations for the shape of the phase kink,
the amplitude and phase shift of plasma oscillation. An approximate
solution can be obtained easily in the strong inductive-coupling
limit where the phase kink renders a step function. This formalism
releases one from a full computer simulation, and can be extended
from 2D to 3D straightforwardly.

The present paper is organized as follows. In Sec.~II, we formulate
the kink state for the (1,0) cavity mode (thus 2D) in a rectangular
mesa presented in Ref.~\cite{Lin08PRL}. The analysis is then
extended to 3D in Sec.~III and the (1,1) cavity mode is analyzed.
The current-voltage curve of the kink state is presented, and the
distribution of electromagnetic field is derived. Section IV is
devoted to analysis for the cylinder geometry. Radiation of THz
electromagnetic waves is addressed under a boundary condition of an
effective impedance in Sec.~V. Discussions are given in Sec. VI
before a brief summary.

\section{1D kink in rectangular mesa}
The coupled sine-Gordon equations for a stack of Josephson junctions
\cite{Tachiki,Sakai93,Kleiner00} take the following form when the
inductive coupling $\zeta$ is large

\begin{equation}\label{csg}
(\partial_x^2+\partial_y^2)P_l=(1-\zeta \Delta^{(2)}) (\sin
P_l+\beta\partial_t P_l +\partial_t^2P_l-J_{\rm{ext}}),
\end{equation}

\noindent where $P_l$ is the total gauge-invariant phase difference
of the $l$-th junction, the lateral directions are scaled by
$\lambda_c$, time by the plasma frequency $\omega_p$, external
current by the critical Josephson current $J_c$,
$\zeta\equiv\lambda_{ab}^2/sD$,
$\beta\equiv4\pi\sigma_c\lambda_c/c\sqrt{\varepsilon_c}$ and
$\Delta^{(2)}Q_l=Q_{l+1}+Q_{l-1}-2Q_{l}$, with $\lambda_{ab}$ and
$\lambda_{c}$ the penetration depths, $s$ ($D$) the thickness of the
superconducting (insulating) layer, $\sigma_c$ the $c$-axis
conductance due to quasi particles, $\varepsilon_c$ the dielectric
constant, and $c$ the light velocity in vacuum
\cite{szlin08,Lin08PRL}. We consider first the system under the
boundary condition $\partial_n P_l=0$ with $n$ the normal to the
sample edges \cite{Lin08PRL,Koshelev08arXiv}.

For the kink solution of the (1,0) cavity mode of a rectangular
mesa, the phase difference $P_l$ is given by \cite{Lin08PRL}

\begin{equation}\label{phase}
P_l(x,t)=\omega t+A\cos(\frac{\pi x}{L_x})\sin(\omega t+\varphi)+
f_{l}P^s(x),
\end{equation}

\noindent where $f_{l}P^s(x)$ is a kink from $0$ to $\pm\pi$ with
the center at $x=L_x/2$. The first linear term is in accordance with
the ac Josephson relation.

In a state characterized by the phase Eq.~(\ref{phase}) with
$f_l=(-1)^l$ or $f_l=(-1)^{[l/2]}$, Eqs.~(\ref{csg}) are decoupled.
From the static parts, the relation of current conservation is
derived \cite{Lin08PRL}:

\begin{equation}\label{current0}
\begin{array}{l}
J_{\rm ext}=\beta\omega-\sin\varphi \int^{L_x}_0\frac{dx}{L_x}
J_{-1}(A g_{10}^{\rm{r}})\cos P^{s},
\end{array}
\end{equation}

\noindent where $g_{10}^{\rm{r}}=\cos(\pi x/L_x)$ and $J_{\nu}$ is
the Bessel function of the first kind. The phase kink $P^s(x)$ is
governed by the differential equation

\begin{equation}\label{kink}
\partial^2_xP^{s} = q\zeta \cos\varphi J_{-1}(A g_{10}^{\rm{r}})\sin P^{s},
\end{equation}

\noindent with $q=4$ for period-2 and $q=2$ for period-4 solutions,
and the boundary condition $\partial_x P^s=0$ at the edges. While
this differential equation has to be solved numerically, to a good
approximation, the phase kink can be described well by
$P^{s}(x)=\phi(x)/2$ with the soliton solution
$\phi(x)=4\arctan[\exp((x-L_x/2)/w)]$. The width of the phase kink
is $w\sim 1/\sqrt{\zeta}\sim 10^{-3}$ for BSCCO, which renders the
kink almost a step function.

From terms associated with $\sin(\omega t)$ and $\cos(\omega t)$
when the phase Eq.~(\ref{phase}) is subject to Eq.~(\ref{csg}), one
obtains

\begin{equation}\label{sinfi}
\begin{array}{l}
\frac{1}{2}A\beta\omega=\\
 \sin\varphi\int^{L_x}_0\frac{dx}{L_x}
g_{10}^{\rm{r}}(J_0(A g_{10}^{\rm{r}})+J_{-2}(A
g_{10}^{\rm{r}}))\cos P^{s},
\end{array}
\end{equation}

\begin{equation}\label{cosfi}
\begin{array}{l}
\frac{1}{2}A\bigl(\omega^2-(\pi/L_x)^2\bigr)=\\
\cos\varphi \int^{L_x}_0\frac{dx}{L_x}g_{10}^{\rm{r}}(J_0(A
g_{10}^{\rm{r}})-J_{-2}(A g_{10}^{\rm{r}}))\cos P^{s}.
\end{array}
\end{equation}

\noindent  With the aids of Eq.~(\ref{sinfi}) and the recursion
relation of the Bessel functions $zJ_{\nu-1}(z)+zJ_{\nu+1}(z)=2\nu
J_{\nu }(z)$, Eq.~(\ref{current0}) is simplified as

\begin{equation}\label{current}
J_{\rm ext}=\beta\omega (1+A^2/4),
\end{equation}

\noindent where the first term clearly counts for the dc ohmic
dissipation, and the second term is caused by the ac voltage
generated by the plasma oscillation, which modifies the otherwise
linear IV curve.

The four equations (\ref{kink})-(\ref{current}) describe the kink
state specified by five quantities, namely $J_{\rm ext}$, $\omega$
(i.e. voltage), $A$, $\varphi$ and $P^s(x)$. Especially, the IV
characteristics can be derived by sweeping the voltage $\omega$.

As the frequency, or equivalently the voltage approaches the cavity
mode $\omega=\pi/L_x$, large plasma oscillations are stimulated and
a large external current is shunted by the Josephson current. Just
at the cavity frequency Eq.~(\ref{cosfi}) is reduced to

\begin{equation}\label{A}
\frac{1}{L_x}\int^{L_x}_0dx g_{10}^{\rm{r}}(J_0(A
g_{10}^{\rm{r}})-J_{-2}(A g_{10}^{\rm{r}})) \cos P^{s}=0.
\end{equation}

\noindent In the limit of strong inductive-coupling, where the phase
kink is well approximated by a step function, Eq.~(\ref{A}) is
easily solved to yield $A=2.331$, and then $\sin\varphi=0.4625$ and
$J_{\rm ext}-\beta\omega=0.2133$ for $\beta=0.02$ and $L_x=0.4$ in
units of $\lambda_c=200\mu m$, typical for the BSCCO single crystal
used in recent experiments.

For comparison, simulations gave $J_{\rm ext}-\beta\omega=0.37$
\cite{Lin08PRL}. Therefore, the present simplified equations give a
semi-quantitative evaluation for the kink state at the cavity modes.

\begin{figure}[t]
\setlength{\unitlength}{1cm}
 \psfig{figure=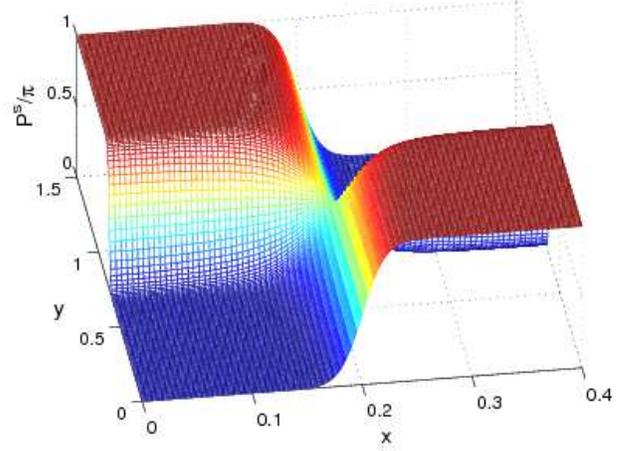,width=\columnwidth}
\caption{\label{f1}(color online). 2D phase kink for the (1,1)
cavity mode of a rectangular mesa obtained by numerical integration
on Eq.~(\ref{kink2D}).}
\end{figure}

\section{2D kink in rectangular mesa}
Now we look for the state of static phase kink of two lateral
dimensions, which corresponds to the (1,1) cavity mode of a
rectangular sample:

\begin{equation}\label{phase2D}
P_l(x,y,t)=\omega t+A\cos(\frac{\pi x}{L_x})\cos(\frac{\pi
y}{L_y})\sin(\omega t+\varphi)+ f_{l}P^s(x,y).
\end{equation}

\noindent The equation for the 2D kink is

\begin{equation}\label{kink2D}
(\partial^2_x+\partial^2_y)P^{s} = q\zeta \cos\varphi J_{-1}(A
g_{11}^{\rm{r}})\sin P^{s},
\end{equation}

\noindent where $g_{11}^{\rm{r}}=\cos(\pi x/L_x)\cos(\pi y/L_y)$
with $\partial_{n}P^s=0$ at edges where $n$ is the normal of the
edges. The solution obtained by numerical integration is depicted in
Fig.~\ref{f1}, with $L_x=0.4$ and $L_y=1.5$ as in experiments
\cite{Ozyuzer07}, and all the prefactors are included into the width
of the kink.

\begin{figure}[t]
\setlength{\unitlength}{1cm}
 \psfig{figure=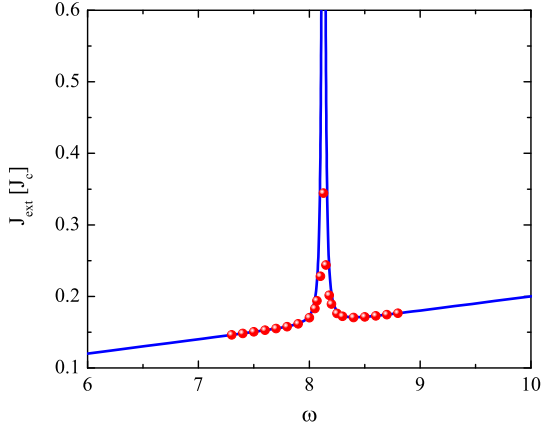,width=\columnwidth}
\caption{\label{f2}(color online). IV curves derived by
Eqs.~(\ref{sinfi2D})-(\ref{current2D}) presuming a step-function
phase kink (red symbols) and by Eq.~(\ref{current2D0}) (blue lines)
for the (1,1) mode of a rectangular mesa with $L_x=0.4$ and
$L_y=1.5$. The cavity frequency is $\omega=8.128$, which corresponds
to 0.62 THz approximately.}
\end{figure}

The remaining three equations are given by

\begin{equation}\label{sinfi2D}
\begin{array}{l}
\frac{1}{4}A\beta\omega= \\
\sin\varphi \int^{L_x}_0\int^{L_y}_0\frac{dxdy}{L_xL_y}
g_{11}^{\rm{r}} (J_0(A g_{11}^{\rm{r}})+J_{-2}(A
g_{11}^{\rm{r}}))\cos P^{s},
\end{array}
\end{equation}

\begin{equation}\label{cosfi2D}
\begin{array}{l}
\frac{1}{4}A\bigl(\omega^2-(\frac{\pi}{L_x})^2-(\frac{\pi}{L_y})^2\bigr)=
\\
\cos\varphi {\int^{L_x}_0\int^{L_y}_0\frac{dxdy}{L_xL_y}
g_{11}^{\rm{r}} (J_0(A g_{11}^{\rm{r}})-J_{-2}(A
g_{11}^{\rm{r}}))\cos P^{s}},
\end{array}
\end{equation}

\begin{equation}\label{current2D}
J_{\rm ext}=\beta\omega (1+A^2/8).
\end{equation}

Approximating $\cos P^{s}$ by the 2D step function (see
Fig.~\ref{f1}), the IV curve is evaluated and displayed in
Fig.~\ref{f2}. An enhancement of the input current appears when the
voltage approaches the cavity mode value
$\omega=\sqrt{(\pi/L_x)^2+(\pi/L_y)^2}$, where the extra energy is
pumped into plasma oscillation.

The way of the system evolving into the resonance state from the
linear ohmic regime can be seen more transparently by neglecting $A$
in the right-hand side of Eqs.~(\ref{sinfi2D}) and (\ref{cosfi2D}).
One then finds

\begin{equation}\label{current2D0}
J_{\rm ext}=\beta\omega
(1+\frac{2{(I_{11}^{\rm{r}})}^2}{(\omega^2-(\pi/L_x)^2-(\pi/L_y)^2)^2+(\beta\omega)^2})
\end{equation}

\noindent with $I_{11}^{\rm{r}}=(1/L_xL_y)\int^{L_x}_0\int^{L_y}_0
dxdy g_{11}^{\rm{r}} \cos P^s$. A similar expression has been
derived for the 1D kink state \cite{Koshelev08arXiv}. The
development of the resonance is captured analytically by the above
expression, although it becomes invalid around the resonating regime
where $A$ is not small any more. In the strong inductive-coupling
limit where the phase kink takes a step function,
$I_{11}^{\rm{r}}=4/\pi^2$.

 As seen in Fig.~\ref{f2}, the two treatments give the same IV
 curves outside the resonance regime where the plasma amplitude is small.
 Usage of Eqs.~(\ref{sinfi2D}) and
 (\ref{cosfi2D}) improves the description of the resonating state over
 Eq.~(\ref{current2D0}).

\begin{figure}[t]
\epsfysize=6.9cm \epsfclipoff \fboxsep=0pt
\setlength{\unitlength}{1cm}
\begin{picture}(8,14.0)(0,0)\epsfysize=6.9cm
\put(-0.5,6.8){\epsffile{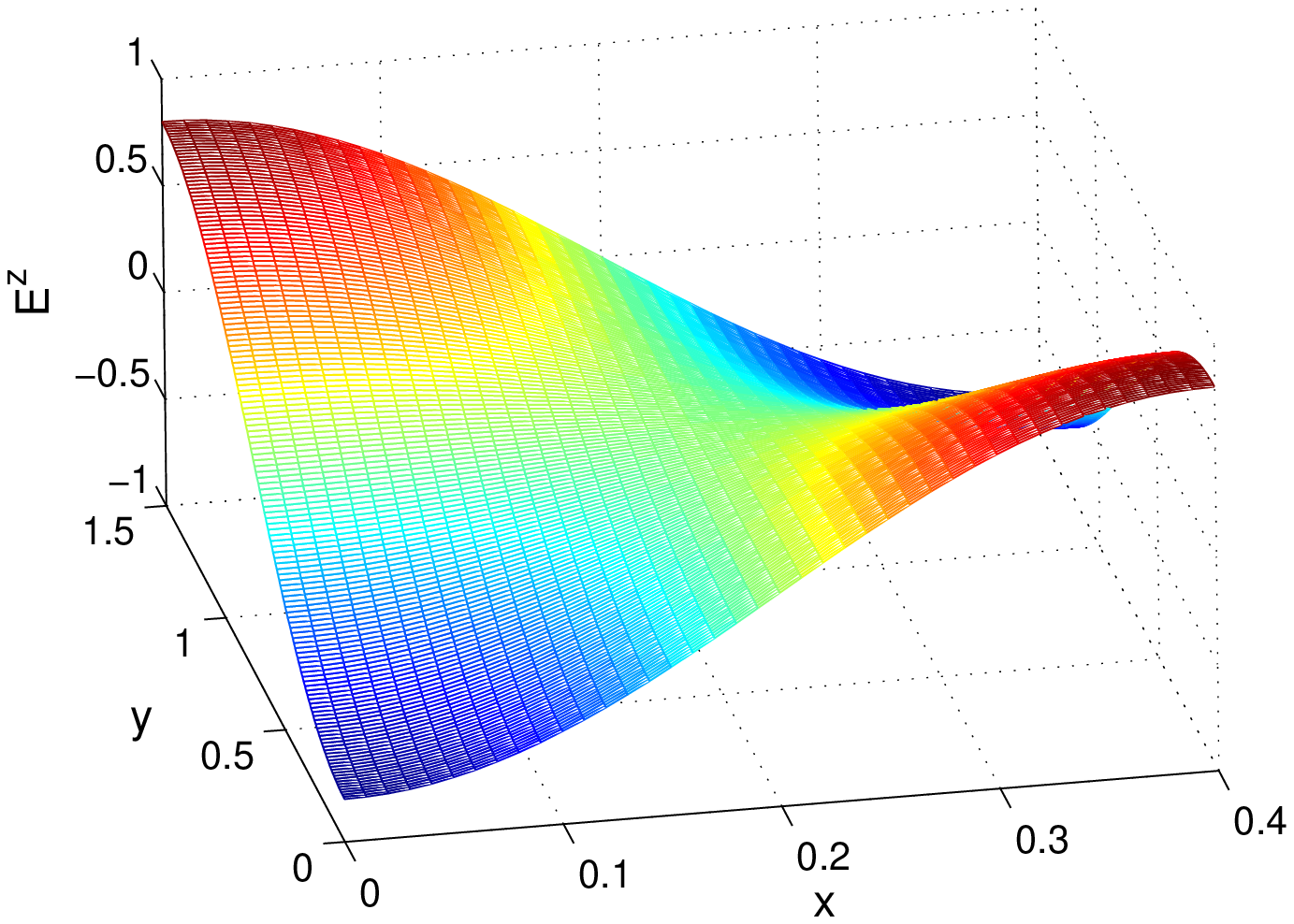}} \epsfysize=6.9cm
\put(-0.5,0.0){\epsffile{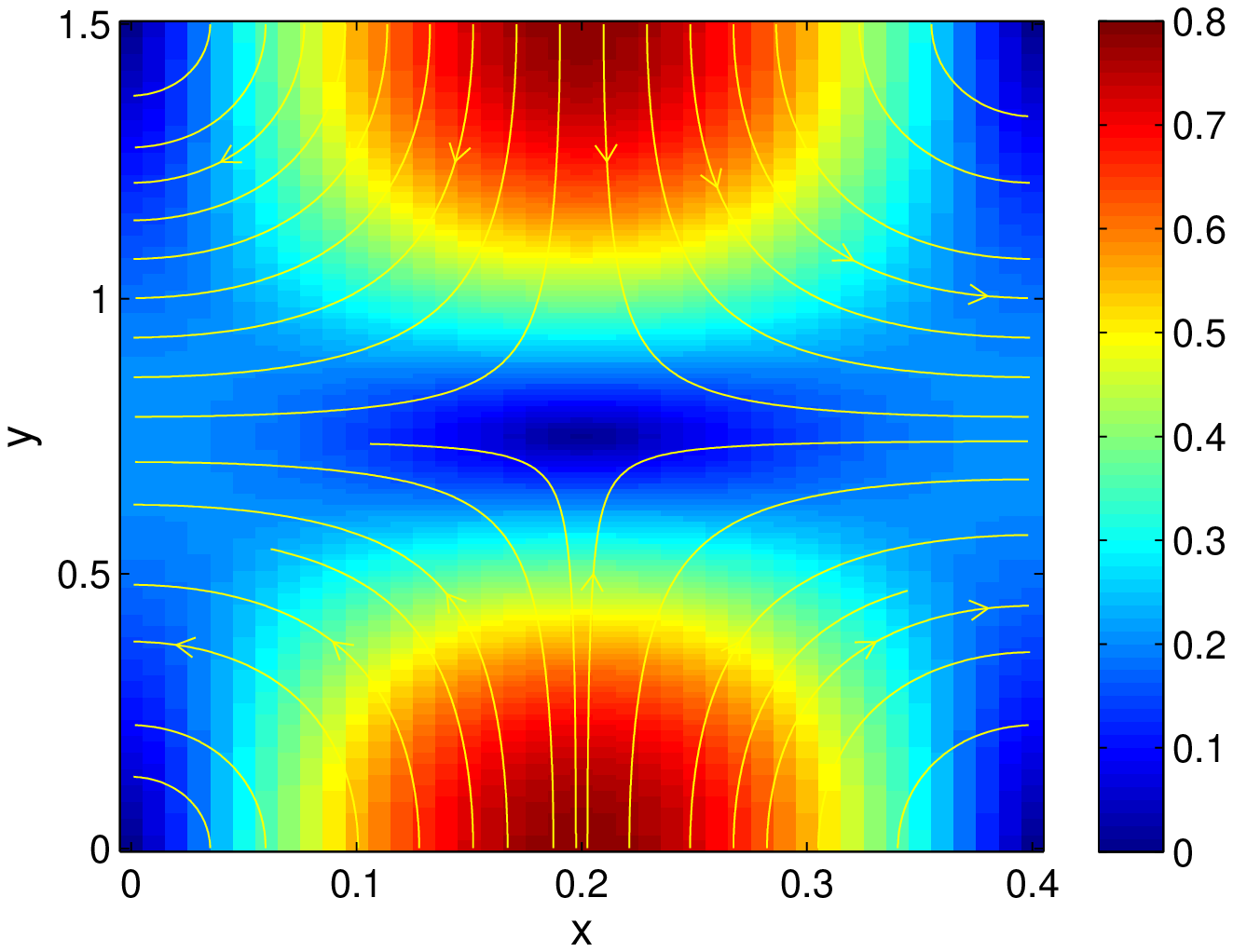}}\put(0.9,12.8){(a)}
\put(0.9,6.6){(b)}
\end{picture}
 \caption{\label{f3}(color online).
Distribution of (a) electric field and (b) magnetic field in for the
(1,1) cavity mode of the rectangular mesa.}
\end{figure}

The electromagnetic fields are given by
$E^z(x,y)=\partial_t\widetilde{P}(x,y)$, $B^x(x,y)\simeq
-\partial_y\widetilde{P}(x,y)$ and $B^y(x,y)\simeq
\partial_x\widetilde{P}(x,y)$, with $\widetilde{P}(x,y)$ the spatial
part of the plasma term in Eq.~(\ref{phase2D}) \cite{Lin08PRL}. As
shown in Fig.~\ref{f3}, the electric field takes the maximal
absolute value at the four corners, whereas the magnetic field is
maximal at the centers of the edges. The magnetic field penetrates
into the system from the two edges along the $x$ direction, and
flows away from the two edges along the $y$ direction, and the
pattern oscillates with time by switching $x$ and $y$. The
$y$-component of magnetic field is larger in absolute value than the
$x$-component since the system is longer in the $y$ direction. Both
the electric and magnetic fields are shielded in the superconducting
sample and thus the central part of the mesa is free of
electromagnetic fields. As discussed in Ref.~\cite{Lin08PRL}, in the
resonance state at cavity modes higher harmonics exist and the
simple symmetry in Fig.~\ref{f3} is modified.

\begin{figure}[t]
\setlength{\unitlength}{1cm}
 \psfig{figure=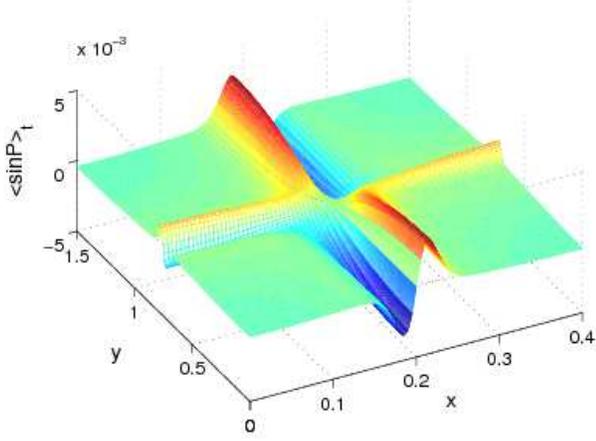,width=\columnwidth}
\caption{\label{f4}(color online). Distribution of static Josephson
current for the (1,1) cavity mode of a rectangular mesa.}
\end{figure}

The static phase kink generates static Josephson currents in
junctions as depicted in Fig.~\ref{f4}. The patterns with swapped
up- and down-ward Josephson currents are stacked periodically along
the $c$ axis according to $f_l=(-1)^l$ or $f_l=(-1)^{[l/2]}$.

\section{Kink state in cylinder geometry}
Now we consider a sample of cylindrical shape and radius $a$. For
the (0,1) mode in the cylinder geometry which is isotropic
azimuthally, the total phase difference is given by

\begin{equation}\label{phasecylinder}
P_l(\rho,t)=\omega t+AJ_0(\frac{v_{01}\rho}{a}) \sin(\omega
t+\varphi)+ f_{l}P^s(\rho),
\end{equation}

\noindent where $v_{01}=3.8317$ is the first zero of $J_1$.

\begin{figure}[t]
\setlength{\unitlength}{1cm}
 \psfig{figure=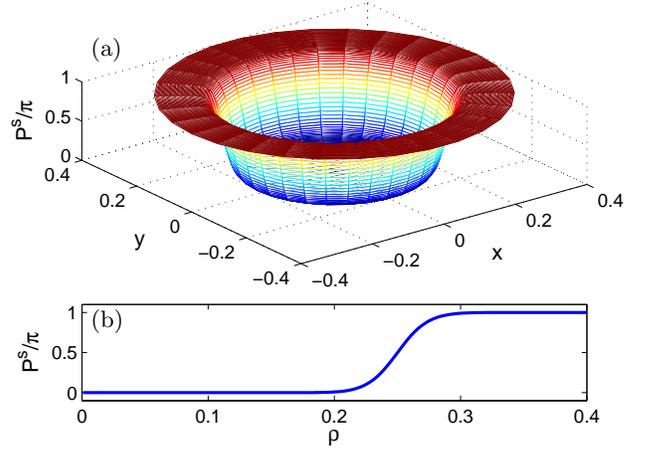,width=\columnwidth}
  \put(-7.4,5.3){(a)} \put(-7.4,1.7){(b)}
\caption{\label{f5}(color online). (a) Phase kink and (b) radial
dependence of phase in a cylinder sample obtained by numerical
integration on Eq.~(\ref{kinkcylinder}).}
\end{figure}

The equation for the static phase kink in cylinder is

\begin{equation}\label{kinkcylinder}
(\partial^2_{\rho}+(1/\rho)\partial_{\rho})P^{s} = q\zeta
\cos\varphi J_{-1}(A g_{01}^{\rm {c}})\sin P^{s},
\end{equation}

\noindent with $g_{01}^{\rm {c}}=J_0(v_{01}\rho/a)$ under the
boundary condition $\partial_{\rho}P^s=0$ at the edge. The solution
obtained by numerical integration is displayed in Fig.~\ref{f5} for
$a=0.4$. The phase crosses $\pi/2$ at $\rho=u_{01}a/v_{01}$ where
$u_{01}=2.4048$ is the first zero of $J_0$.

\begin{figure}[b]
\setlength{\unitlength}{1cm}
 \psfig{figure=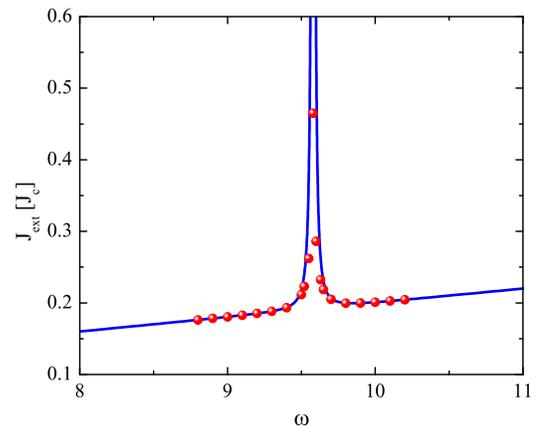,width=\columnwidth}
\caption{\label{f6}(color online). IV curves derived by
Eqs.(\ref{sinficylinder})-(\ref{currentcylinder}) presuming a
step-function phase kink (red symbols) and by
Eq.(\ref{currentcylinder0}) (blue lines) for the (0,1) mode of a
cylinder sample of $a=0.4$. The cavity frequency is $\omega=9.579$,
which corresponds to 0.73 THz approximately.}
\end{figure}

The remaining equations are given by

\begin{equation}\label{sinficylinder}
\begin{array}{l}
\frac{1}{2}A\beta\omega J^2_0(v_{01})=\\
\sin\varphi \int^{a}_0\frac{\rho d\rho}{a^2} g_{01}^{\rm {c}}(J_0(A
g_{01}^{\rm {c}}) + J_{-2}(A g_{01}^{\rm {c}}))\cos P^{s},
\end{array}
\end{equation}

\begin{equation}\label{cosficylinder}
\begin{array}{l}
\frac{1}{2}A\bigl(\omega^2-(v_{01}/a)^2\bigr)J^2_0(v_{01})=\\
\cos\varphi \int^{a}_0\frac{\rho d\rho}{a^2} g_{01}^{\rm {c}}(J_0(A
g_{01}^{\rm {c}}) - J_{-2}(A g_{01}^{\rm {c}}))\cos P^{s},
\end{array}
\end{equation}

\begin{equation}\label{currentcylinder}
J_{\rm ext}=\beta\omega (1+A^2 J^2_0(v_{01})/2),
\end{equation}

\noindent where $J_0(v_{01})=-0.4028$, and the normalization
relation $\int^a_0J^2_0(v_{01}\rho/a)\rho d\rho=a^2J^2_0(v_{01})/2$
has been used. The frequency of the (0,1) cavity mode for the
cylinder geometry is $\omega=v_{01}/a$ which reaches $1$THz when
$a=58.5\rm{\mu m}$.

\begin{figure}[b]
\epsfysize=8cm \epsfclipoff \fboxsep=0pt
\setlength{\unitlength}{1cm}
\begin{picture}(8,14.5)(0,0)\epsfysize=6.4cm
\put(0.0,8.3){\epsffile{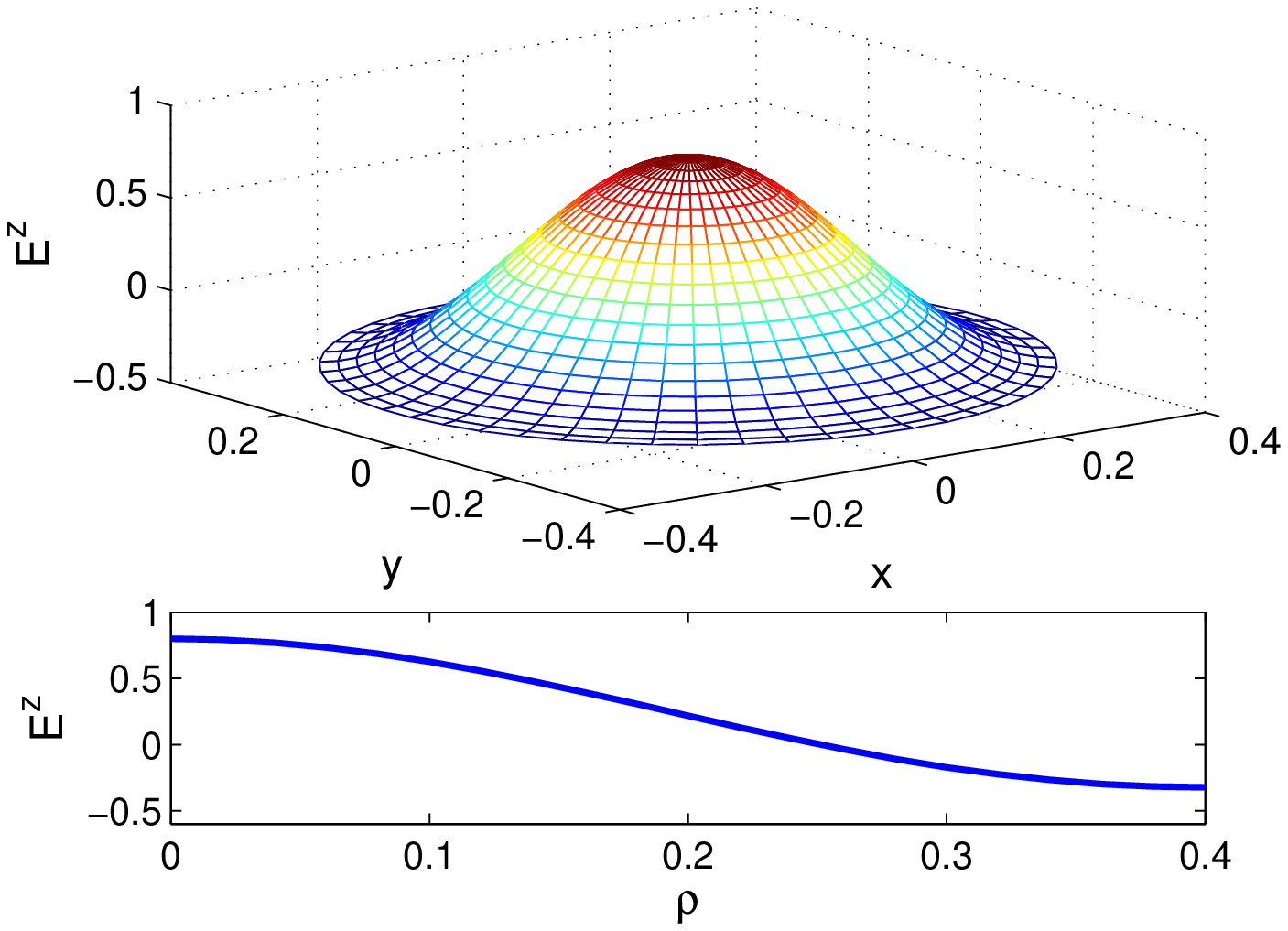}} \epsfysize=8.2cm
\put(0.0,0.0){\epsffile{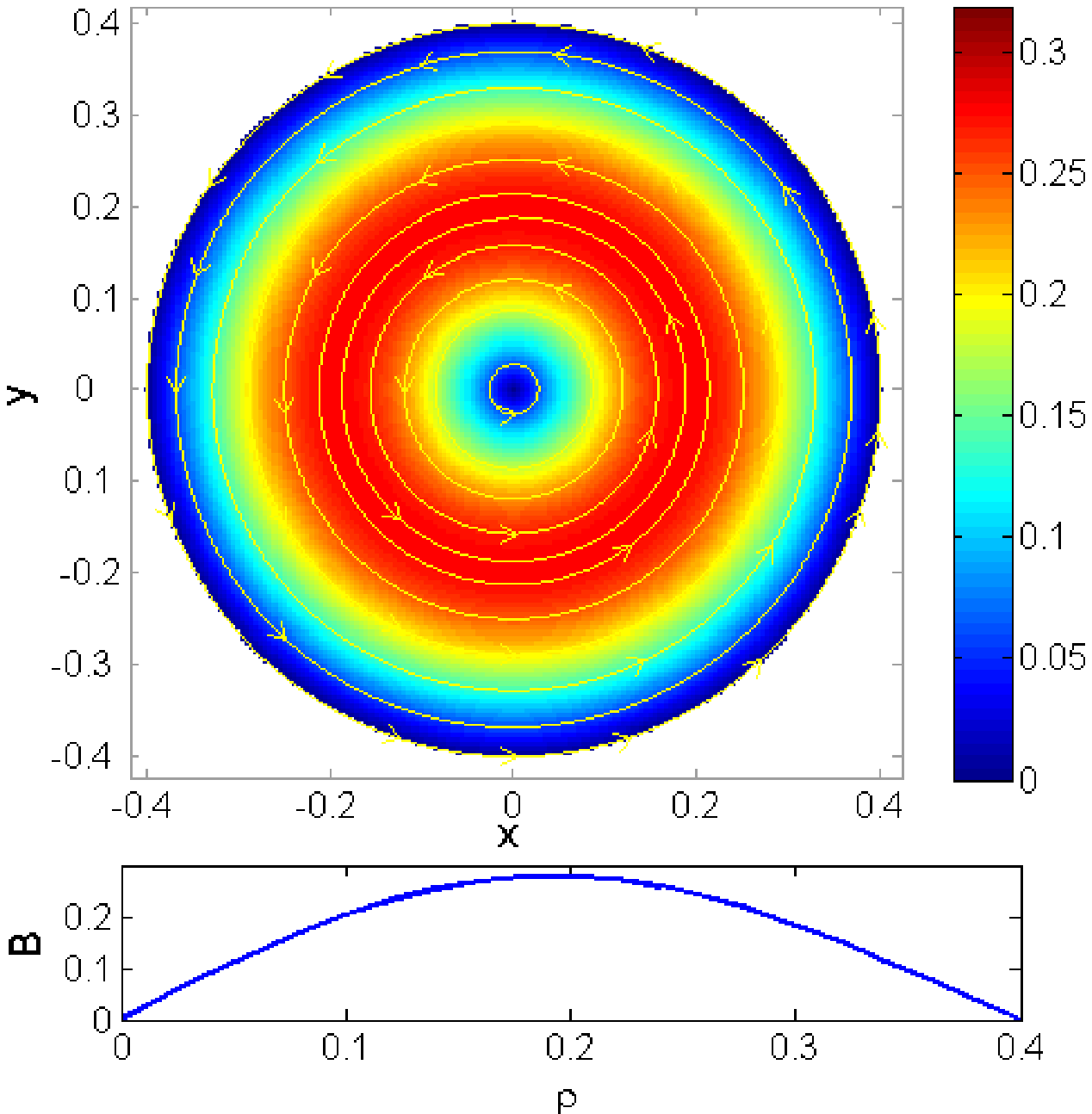}}\put(1.4,13.6){(a)}
\put(1.4,9.2){(b)}\put(1.4,7.7){(c)}\put(1.4,1.4){(d)}
\end{picture}
\caption{\label{f7}(color online). Distribution of electric field
((a) and (b)) and magnetic field ((c) and (d)) in a cylinder
sample.}
\end{figure}

The IV curve is displayed in Fig.~\ref{f6}, taking $\cos P^{s}$ as
the step function. For comparison we also show the IV curve
evaluated by the following expression

\begin{equation}\label{currentcylinder0}
J_{\rm ext}=\beta\omega (
     1+\frac{2{(I_{01}^{\rm {c}})}^2/J^2_0(v_{01})}
            {(\omega^2-(v_{01}/a)^2)^2+(\beta\omega)^2}
                         ),
\end{equation}

\noindent with $I_{01}^{\rm {c}}=(1/a^2)\int^{a}_0\rho d\rho
g_{01}^{\rm {c}}\cos P^s$ obtained in the same way as
Eq.~(\ref{current2D0}). For the step-function phase kink,
$I_{01}^{\rm {c}}=0.1701$. Once again the two evaluations agree with
each other outside the resonance regime.

\begin{figure}[t]
\setlength{\unitlength}{1cm}
 \psfig{figure=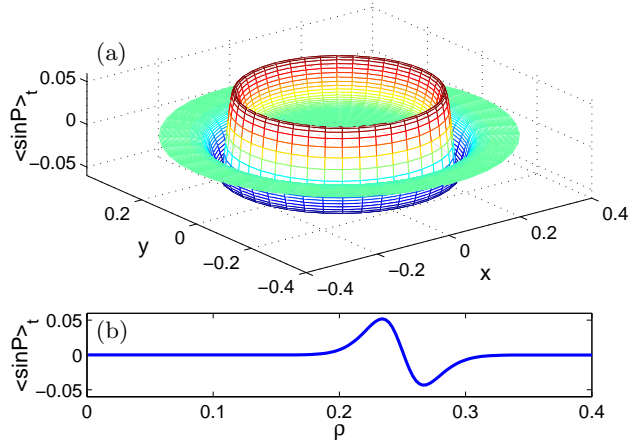,width=\columnwidth}
\put(-7.4,5.2){(a)} \put(-7.4,1.5){(b)}
 \caption{\label{f8}(color
online). (a) Distribution and (b) radial dependence of Josephson
current for the (0,1) mode of a cylinder sample.}
\end{figure}

The electromagnetic fields take the following forms: $E^z=A\omega
J_0(v_{01}\rho/a)$, $B^x=A(v_{01}y/2\rho a)J_1(v_{01}\rho/a)$ and
$B^y=-A(v_{01}x/2\rho a)J_1(v_{01}\rho/a)$. The distribution of
electromagnetic field is shown in Fig.~\ref{f7} which oscillates
with time. The electric field takes the maximum at the center of
cylinder, changes its sign at $\rho=u_{01}a/v_{01}=0.6276a$ and at
the perimeter it reaches $-0.4028$ of the maximum value
(Figs.~\ref{f7}(a) and (b)). The magnetic field is circular, the
amplitude is zero at both the center and edge of the cylinder, and
presumes its maximum at $\sim a/2$ (Figs.~\ref{f7}(c) and (d)). The
distribution of the static Josephson current generated by the static
phase kink is displayed in Fig.~\ref{f8}. The patterns with swapped
up- and down-ward Josephson currents are stacked periodically in the
$c$ axis.

\section{Radiation of energy}
The boundary condition of a single junction subject to radiation was
discussed in literatures
\cite{Langenberg,Bulaevskii06,Bulaevskii06PRL}. There is a
significant mismatch of the impedance between the junctions and
outside space. As in the previous work \cite{Lin08PRL}, we model the
space by an effective impedance $Z$. The effective impedance should
be very large due to the small thickness of the sample compared with
$\lambda_c$ \cite{Koshelev08PRB}. We have confirmed that the cavity
resonance based on the kink state is stable for $|Z|>50$.

The IV characteristics should be modified when the radiation is
present. Since the radiation is governed by the attached impedance,
the IV characteristics can be evaluated simply by an effective
parallel circuit. With a simple dimension counting one finds for a
real impedance $Z$ \cite{Note}

\begin{equation}\label{currentrad}
J_{\rm ext}=\beta\omega +\beta\omega A^2/4 + \omega A^2/(ZL_x),
\end{equation}

\begin{equation}\label{current2Drad}
J_{\rm ext}=\beta\omega + \beta\omega A^2/8 + \omega
A^2(1/L_x+1/L_y)/(2Z),
\end{equation}

\begin{equation}\label{currentcylinderrad}
J_{\rm ext}=\beta\omega + \beta\omega A^2 J^2_0(v_{01})/2 + \omega
A^2 J^2_0(v_{01})/(Za),
\end{equation}

\noindent for the (1,0) mode uniform in the direction of long edge
and the (1,1) mode of the rectangular mesa, and for the (0,1) mode
of the cylindrical mesa, respectively. The density of radiation
energy measured by the Poynting vector then reads

\begin{equation}\label{poynting}
P_{10}^{\rm{r}}= \omega^2 A^2/(2Z),
\end{equation}

\begin{equation}\label{poynting2D}
P_{11}^{\rm{r}}= \omega^2 A^2/(4Z),
\end{equation}

\begin{equation}\label{poyntingcylinder}
P_{01}^{\rm{c}}= \omega^2 A^2 J^2_0(v_{01})/(2Z),
\end{equation}

\noindent for the respective modes.

Presuming the same impedance, we evaluate the ratio between the
Poynting vectors at the corresponding cavity modes when the linear
sample size is the same $2a=L(=L_x=L_y)$:
$P_{01}^{\rm{c}}/P_{10}^{\rm{r}}=3.125$ and
$P_{01}^{\rm{c}}/P_{11}^{\rm{r}}=1.898$, where the maximal
amplitudes $A=2.331$ for (1,0) mode and $2.991$ for (1,1) mode of
the rectangular mesa, and $4.194$ for (0,1) mode of the cylinder
geometry have been used.

There are three factors in determining the Poynting vector: the
maximal plasma amplitude, the factor for the electromagnetic field
at the edge, and the cavity frequency. In cylinders the
electromagnetic fields are isotropic azimuthally, and can make a
full use of the phase kink in pumping dc energy into plasma
oscillations, which results in a large plasma amplitude. The (1,1)
mode in the rectangular mesa is also better than the (1,0) mode
since the former can use a 2D phase kink. While the electric field
is maximal at the edges of the rectangular mesa, the edge electric
field for the cylinder geometry is only $40\%$ percent of its
maximum taken at the center of the cylinder. This suppresses the
radiation power as in Eq.~(\ref{poyntingcylinder}). The cavity
frequency in the cylindrical geometry is superior to the rectangular
geometry by a factor $\sim 2.4$ for the same linear system size.

While the cylinder geometry exhibits a large density of power
emission at the sample surface, the energy is radiated uniformly in
the azimuthal direction. In applications one needs to focus to a
small angle. In contrast, the directivity of the radiation from a
rectangular mesa is helpful for gathering energy efficiently.

As can be seen in Eq.~(\ref{A}) and the counterparts for the other
modes, the plasma amplitude at the cavity resonance does not depend
on the system size. Therefore, we can increase the radiation energy
by adopting smaller samples with larger cavity frequency according
to Eqs.~(\ref{poynting})-(\ref{poyntingcylinder}).

\section{Discussions and Summary}
In the present paper, only the fundamental frequency is kept for the
plasma contribution to the total phase difference. This treatment is
justified when the system is off resonance, where the amplitude of
plasma oscillation is small and higher harmonics can be neglected
safely.

We have checked the validity of this treatment at the cavity
resonance for the cavity mode (1,0) of the rectangular mesa by
comparing the estimates thus obtained with those by accurate
simulations which include automatically all orders of higher
harmonics \cite{Lin08PRL}. We expect that this approximation also
provides reasonable estimates for the (1,1) cavity modes of the
rectangular mesa and for the cylinder geometry.

In the resonating regime, the plasma part is much enhanced, and
higher harmonics appear \cite{Lin08PRL}. The number of equations
governing the kink state is 2$m$+2 when up to the $m$-th harmonics
are covered. Here we provide equations merely including the
fundamental frequency with $m=1$. While the formulas can be extended
easily to higher harmonics, the resultant equations are much more
complex, and take heavy numerics to solve. Discussions on this point
will be reported elsewhere \cite{szlin08Pre}.

In the kink solution, the spatial part of the plasma term is the
eigen function of the Laplace operator in the respective geometry
with zero (or in presence of radiation very small) normal derivative
at the edge, proportional to the magnetic field. In the cylinder
geometry, it gives a cavity frequency $v_{01}/a$ with $a$ the radius
of cylinder and $v_{01}=3.3817$ the first zero of $J_1$. For a state
presuming zero (or very small) electric field at the boundary, the
typical frequency should be $u_{01}/a$, where $u_{01}=2.4048$ is the
first zero of $J_0$. Therefore, detecting the size dependence of the
cavity frequency in the cylinder geometry can tell directly what
state is realized inside the system. Since a rectangular mesa will
give cavity frequency of $\pi/L$ for both of the two cases, the
cylinder geometry is unique in determining the state realized in the
junctions and thus the mechanism of the strong radiation observed
recently in experiments.

The present analysis reveals the way how the stack of Josephson
junctions convert the dc energy to ac electromagnetic radiation, by
formulating out explicitly the oscillation amplitude and the
frequency. The far-field radiation pattern is determined merely by
the spatial part of the plasma term, and is the same as a thin
capacitor subject to an ac driving voltage. Analyses on far-field
radiation patterns for various modes are available in literature
\cite{Leone03IEEE}, and the results can be compared with those
observed for the THz radiations from mesas of BSCCO single crystal.
A detailed discussion will also be given in Ref.~\cite{szlin08Pre}.

To summarize, we have worked out explicitly the kink state in
Josephson junctions of strong inductive coupling in 3D for both
rectangular and cylindrical geometries. A set of equations are
provided which permits one to understand the new state without heavy
numerics. IV characteristics are revealed to be very nonlinear due
to the cavity resonance where the plasma oscillation is much
enhanced. The solution for a cylindrical mesa provides a higher
resonating frequency than that of a square mesa with the same linear
size by a factor of $\sim 2.4$. Investigation on the size dependence
of the resonating frequency in the cylinder geometry can give a
direct evidence for the state realized in the stack of Josephson
junctions, and thus can reveal the mechanism of the strong radiation
discovered recently. Experiments using cylindrical mesas are highly
anticipated.

\section{Acknowledgements}
The authors thank M. Tachiki and K. Kadowaki for discussions. This
work was supported by WPI Initiative on Materials
Nanoarchitectonics, MEXT of Japan and by CREST-JST, and partially by
ITSNEM of CAS.


\begin{thebibliography}{99}
\bibitem{Note} A better treatment is possible and will be reported
elsewhere.
\end{thebibliography}
\end{document}